\documentstyle[11pt,epsfig]{article}
\title{\bf Probabilistic Evolutionary Process: a possible solution to the problem of time in quantum cosmology and
creation from nothing}
\author{N. Khosravi$^{1,2,}$\thanks{email: n-khosravi@sbu.ac.ir, nima@ipm.ir} and H. R. Sepangi$^{1,}$\thanks{email: hr-sepangi@sbu.ac.ir}\\
{\small $^1$Department of Physics, Shahid Beheshti University, Evin,
Tehran 19839, Iran}
\\{\small $^2$School of Physics, Institute for
Research in Fundamental Sciences (IPM), P.O.Box 19395-5531, Tehran,
Iran}}

\begin{document}
\maketitle
\begin{abstract}
We present a method, which we shall call the probabilistic
evolutionary process, based on the probabilistic nature of quantum
theory to offer a possible solution to the problem of time in
quantum cosmology. It offers an alternative for perceiving an arrow
of time which is compatible with the thermodynamical arrow of time
and makes a new interpretation of the FRW universe in vacua which is
consistent with creation of a de Sitter space-time from nothing.
This is a completely  quantum result with no correspondence in
classical cosmology.
\end{abstract}

\maketitle
\section{Introduction}
In this work we show that the probabilistic nature of quantum
mechanics can play an essential role to suggest a possible solution
to the problem of time in quantum cosmology. We have given the
acronym ``Probabilistic Evolutionary Process'' (PEP) to this method.
In addition to the problem of time mentioned above, PEP can account
for the arrow of time too, which is compatible with thermodynamical
arrow of time.  PEP also opens a new window to study quantum
cosmological scenarios such as  the ``creation from nothing"
scenario which has been the focus of attention over the past
decades. Let us then start by focusing on the problem of time.

\subsection{The problem of time}
One of the intriguing notions in theoretical physics is the meaning
of time which plays a crucial role in classical as well as quantum
physics. Over the past decades, a huge amount of effort has been
concentrated on defining exactly what one means by time. In
Newtonian mechanics, time is an external parameter upon which the
evolution of other independent parameters depends. So to describe
the evolution of a system, the parameters of the system are written
in terms of time. However, as we now know, Newtonian mechanics is
only correct when the speeds involve are small compared to the speed
of light or the gravitational fields are weak. We also know that the
fundamental theory is General Relativity (GR). In GR, time is a
coordinate like the others. General relativity is based on the
principle of general covariance which basically states that all
observers must see the same physics. This principle causes GR to
become a gauge theory or, as is required of such theory, invariant
under diffeomorphism transformations. The diffeomorphism invariance
suppresses any manifestation of time in the quantum version of such
theories. This absence of time in diffeomorphism invariant theories
is known as the problem of time, meaning that there is no evolution
in such theories, for a comprehensive review see \cite{isham}. Not
surprisingly, to address this problem in GR and cosmology, a
considerable amount of work has been done over many years
\cite{isham}. In this work we shall offer a new prescription as an
alternative which may address the problem of time in such theories.
This method is based on the probabilistic nature of the wave
functions in quantum mechanics.

The notion of time suffers from another problem as well and that is
the problem of the direction of evolution \cite{arrow}. As is well
known, all the fundamental theories are invariant under time
reversal, namely, $t\rightarrow-t$. This of course means that going
along a specific direction in time is the same as going along the
opposite direction. However, in retrospect, nature seems to take a
preferred direction for time. For example, in thermodynamics the
arrow of time is naturally defined by the second law of
thermodynamics which states that entropy can only increase. The same
is true in observational cosmology in which observations show that
the universe is expanding\footnote{At least we are presently in the
era of an expanding universe.}. A more tangible example is the
psychological arrow of time, that is the ordinary sense of time in
every day life when we make a distinction between the past and
future. Among the above we concentrate on thermodynamical and
cosmological arrows of time. An important question to ask then is:
are these different arrows of time compatible with each other
\cite{arrow}? We will argue that using PEP, one can accommodate an
arrow of time which is compatible with the fundamental
thermodynamical concepts.

\subsection{Creation from nothing}
It would be very interesting if the present universe with all its
immense complications could be shown to have evolved from the
simplest of initial states, namely the vacuum. This is the basic
motivation behind the efforts of those who believe in a unified
theory of everything. The concept of a unified theory led to the
construction of the electro-weak theory and later to the standard
model in particle physics which accounts for the strong
interactions as well. Nevertheless, the intractable gravitational
force is still stubbornly difficult to accommodate into this
scheme. A huge amount of work has been done to combine
gravitational forces with others, but to no avail. The most
promising example is the string theory which has made noticeable
strides towards this goal, but is still far from having a firm
ground to stand on.

One of the approaches to unification is the geometrization of
matter fields. This means that the matter fields are attributed to
and emerge from the geometry of the space-time. The Kaluza-Klein
model \cite{KK} belongs to this category where the appearance of
an extra dimension can play the role of the electromagnetic four
vector potential. On the other hand, in quantum field theory in
curved space-time, it is shown that the expansion of the universe
leads to the creation of matter particles \cite{CST}. In these
models it is often the case that the process of expansion is
introduced by hand and is therefore artificially woven into the
fabrics on which the model is based without any fundamental reason
as to the existence of such processes. Recently, Ambjorn et al.
\cite{loll} have shown that the present accelerating phase of the
universe could have resulted from  quantum dynamical gravity as a
result of the causal dynamical triangulation (CDT).

\section{Probabilistic Evolutionary Process}
To quantize a classical model, the following procedure is commonly
followed. The classical Hamiltonian is written in its corresponding
operator form where, upon quantization, a Schr\"{o}dinger type
equation, i.e. $i\hbar \frac{\partial}{\partial
t}\Psi={\cal{H}}\Psi$, becomes the prevalent dynamical equation from
which the time evolution of the quantum states may be ascertained.
However, in diffeomorphism invariant models the Hamiltonian becomes
a constraint, that is, ${\cal{H}}=0$. These models cannot then
provide for the evolution of the corresponding states and this means
that in these models all the states are stationary. Well known
examples of such models are GR and cosmology. In quantum cosmology,
Schr\"{o}dinger equation becomes the Wheeler-deWitt equation (WD),
${\cal{H}}\Psi=0$. The problem is then arises as to how one can
describe the evolution of the universe, since the universe is not in
a stationary mode as far as the present observational data suggests.
To provide an answer to the question of time, different proposals
have been introduced in different forms, ranging from their
implementation before or after quantization or discarding time
altogether \cite{isham}. However, it seems as if none of these
mechanisms work for simple models such as the FRW metric in vacua.
Since almost all of these proposals introduce a parameter (field) to
represent time, they need at least another parameter to describe its
evolution. In other approaches, one considers the Hamiltonian
constraint itself to find a relation between the parameters
(fields), since one has ${\cal{H}}(a,b)=0$. Therefore, one can write
one parameter in terms of the other, e.g. $a(b)$ and $b(a)$, and
interprets them as the relational behavior of different fields.
These two examples show that the above mechanisms cannot work for
models with only one free parameter.

\begin{figure}[th]
\centerline{\includegraphics[width=10cm]{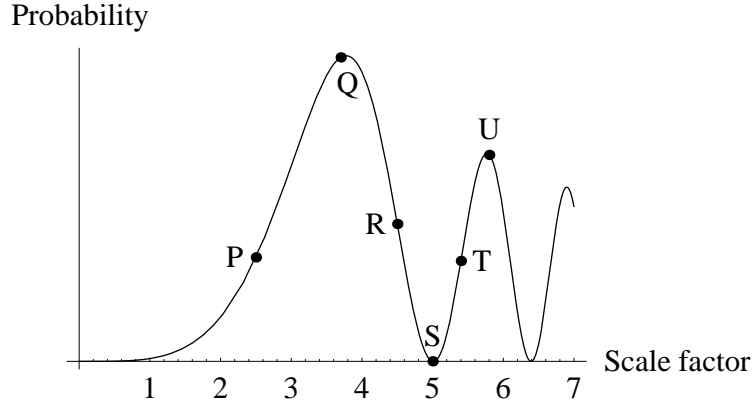}}
\caption{\label{fig1}\footnotesize Figure used to explain the idea
of PEP.}
\end{figure}

In a previous work \cite{nimagrg} we introduced a mechanism based on
the probabilistic structure of quantum systems that can accommodate
systems with only one degree of freedom. In quantum systems the
square or the norm of a state represents the probability, that is,
${\cal{P}}_a=|\Psi(a)|^2$. Now, PEP suggests that the state $\Psi_a$
makes a transition to the state $\Psi_{a+da}$ if their distance,
$da$, is infinitesimal and continuous\footnote{More precisely, the
initial state is around $\Psi_a$ since the ${\cal{P}}_a=|\Psi(a)|^2$
is the probability density.}. The probability of
transition\footnote{This transition probability can play the role of
the speed of transition, i.e. the higher the probability of
transition the larger the speed of transition.} is higher if
${\cal{P}}_{a+da}-{\cal{P}}_{a}$ is larger\footnote{Since there is
no constraint on the positivity of ${\cal{P}}_{a+da}-{\cal{P}}_a$,
PEP can then describe tunnelling processes too. This feature of PEP
was not investigated in \cite{nimagrg}.}. The mechanism for
transition from one state to another is based on utilizing the
effects of a small perturbation\footnote{Certainly, in quantum
cosmology, the universe is considered as one whole
\cite{isham,hall1} and the introduction of an external force is
irrelevant. However, because of the lack of a full theory to
describe the universe, these small external forces are merely used
to afford a better understanding of the discussions presented
here.}$^{,}$\footnote{This perturbation can be caused by the fact
that, for example, the scale factor operator, $A$, does not commute
with the Hamiltonian, ${\cal{H}}$, $[A,{\cal{H}}]\neq 0$. This
physically means that when we restrict the wave function to an
eigenfunction of the Hamiltonian, ${\cal{H}}\Psi(a)=0$, it cannot be
an eigenfunction of $A$ simultaneously i.e. $A\Psi(a)\neq a
\Psi(a)$. So the initial condition $a=a_0$ is not a steady state and
hence it cannot be at $a_0$ and must move away from its initial
value. The rule for such moves are given by PEP.}$^{,}$\footnote{The
origin of the perturbations may be rooted in the von Neumann
approach to quantum mechanics. The perturbation could naturally
arise from the fact that in quantum mechanics one does not know the
position of a state precisely.}. To grasp the features of the
concept of PEP described above in a more clear fashion, we resort to
a simple example given in figure \ref{fig1} which has been borrowed
from the companion paper \cite{nhb}.  Note that the figure is for a
simple model with one degree of freedom (a scale factor). For a more
general model the horizontal axis should be interpreted as the whole
degrees of freedom of the universe such as scale factors, matter
fields and so on.

Let the initial condition be, for example, $a=2.5$, the point $P$
in figure \ref{fig1}. Then PEP states that the system (here the
scale factor or $a$) moves infinitesimally close towards a state
with higher probability and consequently $P$ moves to the right to
reach $Q$, a local maximum and, therefore, stays at $Q$. It means
that the scale factor remains constant as the time
passes\footnote{A perturbation around local maximum is acceptable
as mentioned before.}. We denote this transition by $PEP: P
\rightarrow Q$ or equivalently $P{\buildrel {PEP} \over
\longrightarrow }Q$. Now let the initial condition be the point
$R$. Then we have $R{\buildrel {PEP} \over \longrightarrow }Q$,
and so on. Note that $R{\buildrel {PEP} \over \longrightarrow }S$
is possible but it has much smaller probability due to
$R{\buildrel {PEP} \over \longrightarrow }Q$. We also note that
the transitions $R{\buildrel {PEP} \over \longrightarrow }S$ and
$S{\buildrel {PEP} \over \longrightarrow }T$ can be interpreted as
a tunnelling precess in ordinary quantum mechanics. It means that
PEP can reproduce a tunnelling process but with a very small
probability of occurrence.

The power of addressing time in this fashion is that it is based
on the probabilistic nature of quantum mechanics which is inherent
in the wave function and is deduced here from the WD equation,
hence the notion of time is built on the concept of probability.
In comparison with other models, PEP does not need any extra
treatment like reparameterization by another field or addition of
other fields to realize the desired behavior.

As an evidence for PEP, we have shown that canonical quantization of
cosmology (WD equation) together with PEP is physically equivalent
to the deformed phase-space quantization of cosmology, at least for
the  models discusses in \cite{nhb}. For more thorough understanding
of the discussion at hand, we note that along with the canonical and
path integral approaches to quantization, another procedure also
exists, the so-called deformed phase-space \cite{quantization}. This
approach is well known and much work has been done in this regard
\cite{nhb,nima}. In this kind of quantization, the additional terms
emanating from the deformation of phase space modify the classical
Hamiltonian. These extra terms can either be interpreted as quantum
effects \cite{quantization,nima} or as a quantum potential in the
Bohmian version of quantum mechanics. An important question for
these models is how does the deformed phase-space quantum cosmology
relates to canonical quantum cosmology or the path integral approach
to quantum cosmology? In \cite{nhb} it is shown that based on PEP,
the canonical approach and phase-space deformed approach are
physically equivalent, at least for the models currently under
discussion.

\section{The physics behind PEP}\label{4}
The basis of PEP are on the probabilistic interpretation of
quantum mechanics and more specifically on the probabilistic
interpretation of $|\Psi|^2$, where $\Psi$ is the wave function of
the system under consideration. In a few words, PEP makes a
duality between ``growing in time'' and ``increasing in
probability'', where probability is read from the corresponding
$|\Psi|^2$. To make a convenient description of PEP, we must
insist on a special feature of quantum cosmology which is nothing
but the fact that our system is the universe alone. To be more
clear, we adopt the following two different viewpoints for
describing the universe.

In the first viewpoint, there is an external observer who is outside
the system under discussion. To describe the thermodynamical
properties of that system the observer needs a large number of
copies of the system. These copies can be made by either of the two
following approaches: the observer can make a large number of copies
and observes them simultaneously or the observer has only one copy
of the system and observes it over a long period of time. These two
approaches are standard and equivalent in thermodynamics when
describing a system. Here, the observer can see all the
possibilities with the appropriate weights and therefore can, for
example, normalize the results of the observations to 1, and may
thus  predict the future of the system.

In the second viewpoint, the observer is basically the system itself
and can only change its initial conditions as long as it remains
close to the initial state of the system. The observer cannot see
all the possibilities in the same way as the observer mentioned in
the first viewpoint could, but can only see, as it were, himself and
his neighbors. Now the notion of normalization to 1 is not necessary
and even relevant since in this viewpoint only the relation between
the initial point and its neighbors becomes important. Therefore,
here the notion of probability could be replaced by a more
meaningful one, namely the possibility.

The first viewpoint discussed above is the commonly used method
for describing the behavior of a system since, as an observer
outside the system, we can produce as many copies of the system as
we want and calculate all the appropriate averages. However, this
is not the case when the system under consideration is the
universe itself. There is only one copy of this system and, more
importantly, the observer is internal to the system and not
external. Therefore, mathematically, in assigning $|\Psi|^2$ to
the universe we do not need to normalize the wave function since
as mentioned above, the universe or equivalently the observer can
only see its neighbors. Therefore, it is more convenient to ignore
the notion of probability and change it to possibility  in
describing the second viewpoint. It is worth mentioning that in
general, as is well known, because of the inner product problem in
quantum cosmology, the notion of probability is not well defined.
Such a notion however, becomes redundant in PEP and is replaced by
the notion of possibility which, in spite of the inner product
problem, is well defined in the present context and can be used
unambiguously.

To make the discussion more clear, consider a particle moving under
the influence of a potential of a certain field. From the point of
view of the particle, or an observer moving with the particle, it
can, in principle, move infinitesimally close to any of its
physically allowed neighboring points, but it chooses a point with a
lower potential since it experiences the force $\vec{F}\propto
-\nabla V$, where $\vec{F}$ and $V$ are the force and its
corresponding potential respectively. In summary, a standard
particle moves without any prior knowledge of the properties of the
far points (points except those that are in its neighborhood) and as
a result, the particle would end up in a local minimum of the
potential and not necessarily in a global minimum. This is much the
same as the behavior of the universe taken as the system, discussed
above. In quantum cosmology, the degrees of freedom of the
minisuperspace play the role of position in the above example and
the function $|\Psi|^2$ plays the role of the potential. We shall
present an extended discussion on PEP in the Conclusions section.

\section{Thermodynamical arrow of time and PEP}\label{2}
The root of thermodynamical arrow of time is in the second law of
thermodynamics (SLT), stating that ``entropy is not decreasing", or,
putting in mathematical form $\bigtriangleup S\geq0$, where $S$ is
the entropy of a closed system. Note that the entropy of a system,
like its energy, becomes meaningful only if compared to a defined
standard or another system. The second law of thermodynamics opens
the door to an important physical controversy, namely time reversal
since any macroscopic system would evolve to a more disordered
state, starting from an initially ordered sate. To account for this
transition one can count all the possible micro states of the system
and calculate the corresponding entropy as
\begin{eqnarray}\label{entropy}
S=k_B\log N
\end{eqnarray}
where $k_B$ is the Boltzman constant and $N$ is the number of all
the possible different microstates for a definite macrostate.
Explicitly,  it means that for a given macroscopic system with a
finite number of degrees of freedom, finite volume and finite
temperature, the number of allowable microstates is $N$. This
shows the high degree of correlation between thermodynamics and
combinatorial arithmetics of micro-structures. Here the similarity
of SLT and PEP becomes clear since they have a common base, the
microstates possibilities.

This monotonic behavior of  entropy is very convenient for a
definition of time. It means that time is a monotonic function of
the entropy which can mathematically be stated as $\triangle
t=t_f-t_i=f (\triangle S=S_f-S_i)$ in which $f$ is a monotonically
increasing function and $S_f$ and $S_i$ are the entropy of the final
and initial states at times $t_f$ and $t_i$ respectively. Note that
the function $f$ is free of any constraint except the monotonically
increasing behavior and so its form is unknown at least up to the
uncertainty in our knowledge at present.

An important problem here is that when we speak of the universe as a
whole what becomes of the meaning of microstates? We have only one
macrostate, the universe! Here, we just assume a duality between PEP
and SLT due to their common sensitive responses to possibilities.
One suggestion is that we can interpret isotropicity and homogeneity
constraints in quantum cosmology, as macroscopic
constraints\footnote{Also, weaker or stronger constraints. This
viewpoint is not all that strange since really the isotropicity and
homogeneity are macroscopic symmetries but are broken in microscopic
regimes, e.g. in the Milky way.}. So we can interpret the
constraints as defining the macroscopic states i.e. the macroscopic
structure is fixed by the constraints like temperature, pressure and
the number of particles in the thermodynamical system which would
define the macroscopic structure of the system. The resulting wave
function that satisfies the macroscopic constraint then shows the
possibilities of microstates which satisfy the macroscopic
constraints and also the weight of any one of these possibilities
with respect to the others. In this view the relation between PEP
and SLT becomes more clear.

\section{Creation from nothing}\label{no}
Here we study a simple FRW model to show how PEP can predict
non-trivial (non-vacuum) solutions from trivial (vacuum) ones. Let
us take the FRW metric with zero curvature
\begin{eqnarray}\label{frw}
ds^2=-N^2(t)dt^2+{a^2(t)}(dx^2+dy^2+dz^2),
\end{eqnarray}
where $N(t)$ and $a(t)$ are the lapse function and scale factor
respectively. The corresponding action becomes
\begin{eqnarray}\label{lagrangian}
{\cal{L}}&=&\sqrt{-g}(R[g]-V(a))\nonumber\\
&=&-6N^{-1}a\dot{a}^2- Na^3V(a),
\end{eqnarray}
where $V(a)$ is related to an arbitrary matter field and the total
derivative terms are ignored in the second line. The corresponding
Hamiltonian becomes
\begin{eqnarray}\label{hamiltonian1}
{\cal{H}}_0&=&\frac{1}{24}Na^{-1}p_a^2- Na^3V(a).
\end{eqnarray}
Since the momentum conjugate to $N(t)$ does not appear in the above
Hamiltonian it is a primary constraint. Therefore, to obtain the
full equations of motion we shall work with the Dirac Hamiltonian
which is more appropriate
\begin{eqnarray}\label{hamiltonian1}
{\cal{H}}&=&\frac{1}{24}Na^{-1}p_a^2- Na^3V(a)+\lambda \pi,
\end{eqnarray}
where $\pi$ is the momentum conjugate to $N(t)$ which is added
through a Lagrange multiplier, $\lambda$, as a primary constraint.
The corresponding equations of motion are
\begin{eqnarray}\label{eqofmotion}
\dot{a}&=&\left\{a,{\cal{H}}\right\}=\frac{1}{12}Na^{-1}p_a,\nonumber\\
\dot{p_a}&=&\left\{p_a,{\cal{H}}\right\}=\frac{1}{24}Na^{-2}p_a^2+3
N a^2V(a)+N a^3 V'(a),\nonumber\\
\dot{N}&=&\left\{N,{\cal{H}}\right\}=\lambda,\nonumber\\
\dot{\pi}&=&\left\{\pi,{\cal{H}}\right\}=-\frac{1}{24}a^{-1}p_a^2+a^3
V(a),\label{constraint}
\end{eqnarray}
where a prime represents differentiation with respect to the
argument. To preserve the primary constraint, $\pi=0$, at all
times the secondary constraint must be satisfied naturally i.e.
$\dot{\pi}=0$. Due to the latter constraint and the above
equations, the equation of motion in the comoving gauge $N(t)=1$
becomes
\begin{eqnarray}\label{eqofmotion1}
\dot{a}=\sqrt{\frac{1}{6} a^2 V(a)},\label{1equation}
\end{eqnarray}
which is the familiar Friedman equation. The above equation has
been solved for different kinds of matter which are represented by
$V(a)$, such as radiation, dust and the cosmological constant. It
is obvious that the above equation for vacuum, $V(a)=0$, reduces
to the trivial Minkowskian metric. It means that classical general
relativity predicts only the trivial solution for an isotropic and
homogeneous space-time.

Now let us study the above simple model in the quantum regime. To
quantize the model we follow the Dirac approach to get the
Wheeler-DeWitt equation as $\hat{{\cal{H}}}\Psi(a)=0$ which for
our model is
\begin{eqnarray}\label{WD}
\hat{{\cal{H}}}\Psi(a)=\left[\frac{1}{24}p_aa^{-1}p_a-a^3V(a)\right]\Psi(a)=0,
\end{eqnarray}
where a certain ordering is assumed and $[a,p_a]=i$ ($\hbar=1$).
In the $a$-representation, the above equation transforms to the
following differential equation
\begin{eqnarray}\label{diffWD}
\partial_a^2\Psi(a)-a^{-1}\partial_a\Psi(a)+{24}
a^4V(a) \Psi(a)=0.
\end{eqnarray}
To compare quantum solutions with the classical ones, we restrict
ourselves to the vacuum case $V(a)=0$. The solution becomes
\begin{eqnarray}\label{vacsol}
\Psi(a)\propto \left\{%
\begin{array}{lll}
c_1,\\
\\
c_2 a^2, \\
\end{array}%
\right.
\end{eqnarray}
where $c_1$ and $c_2$ are integration constants. Here we make our
interpretation using PEP to describe the above solutions.
\begin{figure}
\begin{tabular}{ccc} \epsfig{figure=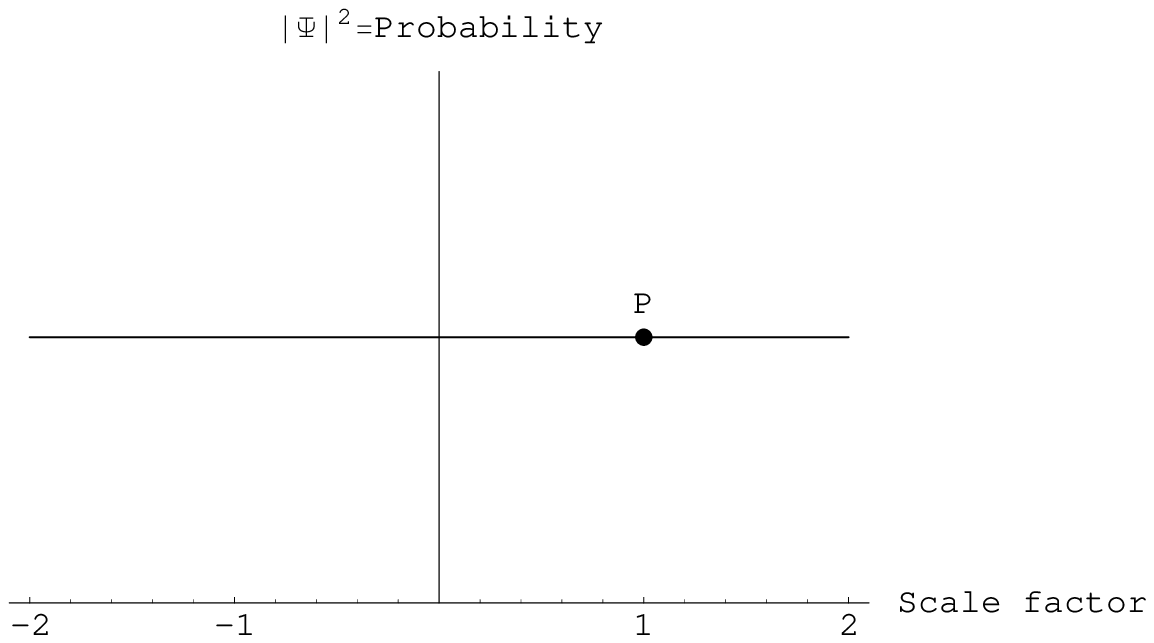,width=7cm}
\hspace{1cm} \epsfig{figure=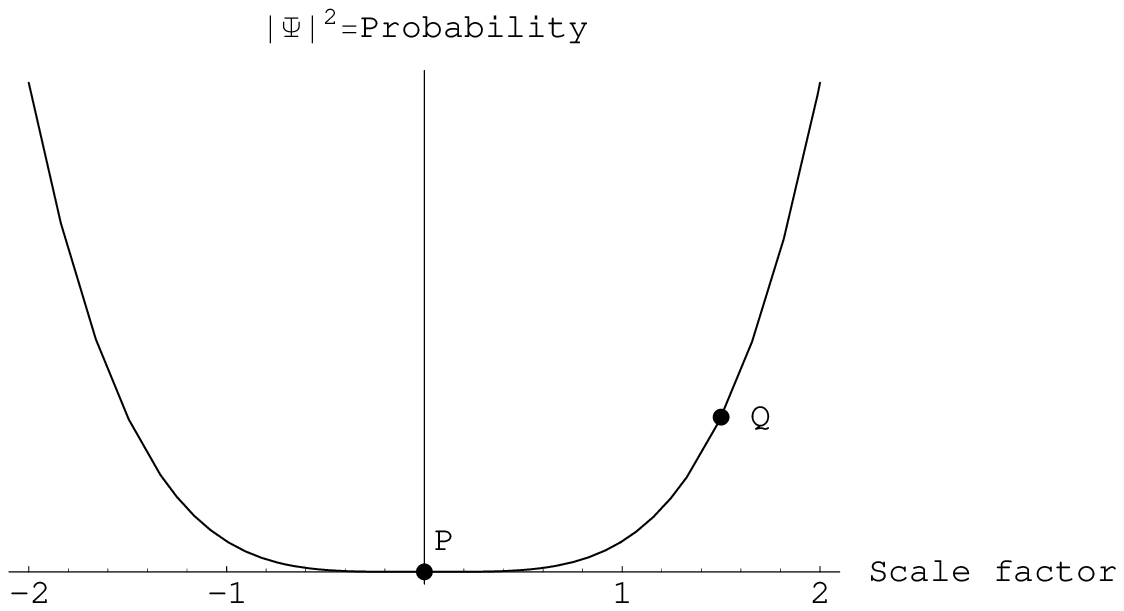,width=7cm}
\end{tabular}
\caption{\label{mi}\footnotesize The left figure shows the trivial
case and the right one the non trivial (quantum) case.}
\end{figure}
\subsection{First case}
For the first solution if one chooses an initial condition for the
scale factor, it remains in this initial condition since the norm
of the scale factor is a constant i.e. $P{\buildrel {PEP} \over
\longrightarrow }P$, as the left plot in figure \ref{mi} shows. It
means that the scale factor is a constant which is the trivial
Minkowskian solution similar to the classical solution.

\subsection{Second case}
This case is our main result and has no counterpart in the classical
case, noting that all the results here are of quantum nature. The
norm of the scale factor predicts an unfinished expansion for the
scale factor due to PEP i.e. $P{\buildrel {PEP} \over
\longrightarrow }Q$, the right plot in figure \ref{mi}. Note that
this behavior does not have a classical counterpart. Physically, it
means that quantum effects cause an expansion even for vacuum. This
prediction is important for the unification of all forces. It also
means that if the initial state is the vacuum, the quantum effects
cause the expansion and the expansion in turn creates particles. In
summary, the present state of the universe with all matter fields is
a result of the vacuum state. This is only a result of the dynamical
interpretation of quantum cosmological solutions by PEP.

Such interpretation of the wave function makes the usual WD equation
comparable with other, more complicated models like the Causal
Dynamical Triangulation method (CDT) for quantizing general
relativity. In \cite{loll} it is shown that the present de Sitter
phase of the universe can be reached from a vacuum initial condition
due to the evolution rules laid down by CDT. If one believes in the
results of \cite{loll}, some approximation can then be presented
using the PEP. For example, in \cite{loll} the resulting universe
has an exponential behavior in time, so a rough calculation shows,
using (\ref{vacsol})
\begin{eqnarray}\label{peploll}
\left\{%
\begin{array}{lll}
a(t)=e^{\sqrt{\frac{\Lambda}{3}}t}\\
\hspace{3cm}\Rightarrow\hspace{1cm} \mbox{time}= \frac{1}{L} \log(p/p_0),\\
p=c_2^2a^4 \\
\end{array}%
\right.
\end{eqnarray}
where $L=\sqrt{\frac{3}{16 \Lambda}}$ and $p_0$ is an arbitrary
constant. Note that in the PEP viewpoint we do not have any
cosmological constant, and therefore to make it consistent we must
rewrite the multiplier with an appropriate constant which is defined
in the model. Since in quantum cosmology  $c$, $G$ and $\hbar$ are
defined, one must write $L$ as a function of these constants in a
such a way that $L$ has the dimension of 1/time. So it is natural to
choose $L=t_p^{-1}$ and therefore\footnote{It would be interesting
to observe that if one believes in the relation between entropy
(\ref{entropy}) and time (\ref{peploll}), then a relation between
the Boltzmann constant and Planck scale becomes natural.}
\begin{eqnarray}\label{peptime}
t=\sqrt{\frac{\hbar G}{c^5}} \log{p/p_0}.
\end{eqnarray}
Is the above result exact? Not really, even if we believe that the
quantum vacuum will lead to an exponential behavior for the
universe. Since in the present epoch the behavior of the scale
factor is believed to be of a power law type, having a different
type for earlier times, it then makes sense to regard it as possible
that the form of the relation between time and probability must be
changed at least for small scale factors.

Note that as was mentioned, the above results are direct
consequences of quantum geometry. In summary, the vacuum state of
quantum geometry may lead to a non-vacuum state in a classical
framework. This result is the goal of all physicists who pursue
the notion of unification.

Now, suppose the initial scale factor is zero, $a(0)=0$, point $P$
in figure 2. In this case the 3-geometry becomes the 0-geometry
and is in an unstable equilibrium. The universe exits from this
point due to PEP and the above discussions become relevant. The
point here is that one can make a similarity between the initial
0-geometry, point $P$, and the Vilenkin's creation from nothing
such that ``nothing refers to the absence of not only matter but
also space and time'' \cite{vil}. We note that in our simple
model, for the initial point $P$ we have neither space, since it
is the 0-geometry, nor matter, as assumed by imposing the
condition $V(a)=0$.

\section{Conclusions}
We have introduced a method to interpret the evolution of the wave
function of the universe using the probabilistic evolutionary
process (PEP). PEP is based on the probabilistic interpretation of
quantum mechanics. The PEP's rule is that the universe can evolve to
a state in its neighborhood if the latter is more probable. We have
shown that this kind of interpretation of the wave function, in
addition to suggesting a possible solution to the problem of time in
quantum cosmology, makes a definition of an arrow of time possible.
The PEP's arrow of time coincides with the thermodynamical one due
to the representation of the latter  by microstate possibilities
(probabilities). In a companion paper \cite{nhb} we have shown that
the prediction of canonical quantum cosmology with PEP is equivalent
to deformed phase space quantum cosmology. This feature can be
interpreted as an evidence for PEP even if this correspondence is
true only for some models. Finally, we have shown that PEP predicts
a nontrivial (e.g. de Sitter) solution from a trivial (vacuum)
quantum state, that is, creation from nothing. One may extend this
method to more complicated models, but even an example as simple as
the one presented in this paper results in interesting  and non
trivial features, namely a possible resolution of question of time
in quantum cosmology and  a mechanism for creation from nothing. Let
us present a quote from \cite{loll} which is particularly relevant
to our discussion here, ``to show that the physical space time
surrounding us can be derived from some fundamental,
quantum-dynamical principle is, a holy grail of theoretical
physics".

Since the PEP is in its first steps of development, there naturally
arises some questions, e.g. what are the equations governing the
dynamics of transition from a low probability states to a higher
one, or what is the correspondence between PEP and semi-classical
wave functions etc. As for the first question, since such a
transition is related to a change in the entropy, the natural choice
to describe the dynamics of the transition resides in the dynamics
of the increasing entropy in non-equilibrium statistical mechanics.
Now, it is commonly known that the evolution of the entropy depends
on the microscopic structure of the macroscopic system under
consideration. As was mentioned above, we may imagine the scale
factor as being a macroscopic quantity so that its evolution would
depend on the microscopic degrees of freedom of the system, namely
the universe. The coarse graining structure of the space time
considered in the literature \cite{quantum gravity} is an important
example relevant to the present discussion. Any further discussion
relating to this matter should naturally await the emergence of a
full quantum theory of gravity. As far as the second question is
concerned, namely the semi-classical wave function, we will see that
the approach is not relevant to our example and results. The first
step in establishing the classical-quantum correspondence is the
decomposition of the wave function according to, $\Psi=R e^{i S}$.
It has been mentioned that in this approach if $S=0$ then one cannot
work in an appropriate manner since the method breaks down
\cite{kiefer}. This is exactly what we encounter here since in our
toy model the absence of any potential term in the Lagrangian causes
the vanishing of $S$. This poses no contradiction to our
interpretation  of the second equation in (\ref{vacsol}) since we
interpret it as a pure quantum result without any counterpart in
classical cosmology.

Finally, an interesting point to note is that  the time variable in
the previous sections (specially in section \ref{no}) is a
coordinate (gauge) variable whereas the entropy seems to be a
quantity which is independent of the observer. The question then
arises as to how such a gauge dependent variable, that is time, can
be related to entropy. This should cause no alarm here since the
relation between the time coordinate and number of possibilities has
some roots in the notion of entropy and SLT. In  section \ref{2}, we
introduced a function $f$ which establishes the correspondence
between time and entropy in such a way as to make the former a
monotonically increasing function of the latter  without any further
constraint. However, the above discussion results in an additional
constraint on this function. Since such a function relates the
number of possibilities (entropy) to time or, a gauge independent
quantity to a gauge dependent quantity, it has to be a gauge
dependent function.

\vspace{10mm}\noindent\\
{\bf Acknowledgments}\vspace{2mm}\noindent\\ We would like to
thank S. S. Gousheh, E. Khajeh and B. Vakili for useful
discussions.

\end{document}